\documentclass[conference]{IEEEtran}
\IEEEoverridecommandlockouts
\usepackage{cite}
\usepackage{amsmath,amssymb,amsfonts}
\usepackage{algorithmic}
\usepackage{graphicx}
\usepackage{textcomp}
\usepackage{xcolor}
\usepackage{url}
\usepackage{enumitem}
\def\BibTeX{{\rm B\kern-.05em{\sc i\kern-.025em b}\kern-.08em
    T\kern-.1667em\lower.7ex\hbox{E}\kern-.125emX}}
\begin{document}

\title{TraceView: Interactive Visualization of Agentic Program Repair Trajectories}

\author{
\IEEEauthorblockN{Amirali Sajadi\IEEEauthorrefmark{1}, Tu Nguyen\IEEEauthorrefmark{1}, Kimmie Huynh\IEEEauthorrefmark{1}, Esteban Parra\IEEEauthorrefmark{2}, Preetha Chatterjee\IEEEauthorrefmark{1}}
\IEEEauthorblockA{\IEEEauthorrefmark{1}Drexel University, Philadelphia, PA, USA\\
\{amirali.sajadi, jn866, kh3472, preetha.chatterjee\}@drexel.edu}
\IEEEauthorblockA{\IEEEauthorrefmark{2}Belmont University, Nashville, TN, USA\\
esteban.parrarodriguez@belmont.edu}
}

\maketitle

\begin{abstract}
LLM-based automated program repair (APR) agents generate patches to fix software bugs with minimal human intervention. These agents often produce long trajectories of reasoning, tool use, and feedback to produce candidate patches. Final patch outcomes show whether a repair attempt succeeded or failed, but they do not show how the agent reached that outcome, or where the process became repetitive or misaligned with the task. This makes agentic repair failures difficult to diagnose, reproduce, and prevent. To help developers address these challenges, we present \textit{TraceView}, an interactive tool for labeling and visualizing repair trajectories from APR systems. TraceView organizes raw and pre-labeled agentic runs with \textit{Thought}, \textit{Action}, and \textit{Result} components to support semantic relation labeling and diagnosis, and renders the resulting trajectory as graph views. Furthermore, TraceView provides relation filters, patch outcome summaries, metrics, and node-level evidence panels to help users inspect how \textit{reasoning,} \textit{actions}, and \textit{feedback} connect across the various steps of an agentic repair attempt. We evaluate TraceView with five researchers through a survey-based user study. Participants reported that TraceView made trajectories easier to scan and that its overview-to-detail workflow helped them better understand repair behavior. The TraceView source code is available at \url{https://github.com/SOAR-Lab/agent-traj-visualization}. A screencast of TraceView is available at \url{https://youtu.be/9ZCh7Ifj2AQ}.
\end{abstract}


\section{Introduction}

LLM-based software engineering agents are increasingly used for APR, issue resolution, and test-driven patch generation~\cite{ehsani2025characteristics,  ehsani2025hierarchical, 10.1145/3650212.3680323, 10.1109/ICSE48619.2023.00128}. Systems such as AutoCodeRover and SWE-agent show that agents can search repositories, inspect code, edit files, and run tests while attempting to generate patches that could fix bugs in software systems~\cite{zhang2024autocoderover, yang2024sweagent}. However, beyond the functionality of the final output, a passing patch does not explain which evidence the agent used, and a failing patch does not show whether the agent localized the bug, ignored feedback, repeated mistakes, or diverged from the task.

Trajectory analysis provides a way to inspect the repair process taken by an agentic APR system \cite{bouzenia2025understandingsoftwareengineeringagents}. APR agent logs often consist of recurring \textit{Thought}, \textit{Action}, and \textit{Result} (TAR) components~\cite{bouzenia2025understandingsoftwareengineeringagents}. A \textit{thought} records the agent's reasoning or plan, an \textit{action} records an operation such as searching files, inspecting code, editing code, or running tests, and a \textit{result} records environmental feedback from the repository, tools, or tests. Beyond these components, recent work has shown the value of labeling the relations in agent trajectories~\cite{Liu_2026, abdollahi2025demystifyingerrorsllmreasoning, bouzenia2025repairagent}. These relation labels describe how one element connects to another, for example, whether a thought leads to an action, a result supports a later decision, an action repeats an earlier one, or feedback has no clear influence on the next step. Such labels are useful because they show where the agent's reasoning, actions, and feedback remain aligned, and where the repair process starts to loop, or diverge from task focus. They can also provide targeted course-correction guidance to nudge the agent back to a productive path~\cite{nanda2026winkrecoveringmisbehaviorscoding}. 

Prior work also shows that producing these relation labels for agent trajectories is costly. For instance, Bouzenia and Pradel manually labeled 14K pairs of TAR components across 120 APR agent trajectories, with the initial annotation requiring months of work~\cite{bouzenia2025understandingsoftwareengineeringagents}. This effort illustrates both the value of relational trajectory analysis and the need for tool support. Without an interface for creating, filtering, and inspecting relation labels, researchers must move between raw logs and spreadsheets, %and ad hoc scripts, 
making it difficult to reuse annotations or connect them back to the evidence in the original trace. This challenge is also consistent with broader principles for interactive analysis~\cite{5290695, 545307}. Users should be able to start with an overview of the repair workflow, narrow the graph to relation types such as repetition, divergence, or no influence, and then inspect the exact thought, action, result, and source text behind each labeled connection.
%users need to move from an overview of the repair workflow to filtering relations such as repetition, divergence, or no influence, to inspecting the exact thought, action, result, and source text behind each labeled connection.

%Visualization research provides a useful foundation for this problem. Addressing these challenges requires interactive tool support for navigating and interpreting agent trajectories. Munzner's nested model emphasizes aligning the domain problem, data abstraction, visual representation, and user operations~\cite{5290695}. Shneiderman's visual information seeking framework similarly emphasizes moving from an overview to filtering and then to details on demand~\cite{545307}. These ideas fit APR trajectories because users need to scan the repair workflow at a high level, filter for relations such as repetition, divergence, or no influence, and inspect the exact thought, action, result, and source text behind each labeled connection.

Recent agent workflow analysis tools demonstrate the value of structuring long execution or reasoning traces before visualization. For example, Anteater visualizes program execution values in context~\cite{faust2024anteaterinteractivevisualizationprogram}; AgentLens organizes agent events into hierarchical behaviors and sub behaviors~\cite{lu2024agentlensvisualanalysisagent}; Agent Trajectory Explorer supports visualization and feedback over agent trajectories~\cite{10.1609/aaai.v39i28.35350}; ReTrace summarizes long reasoning traces through complementary phase based views~\cite{felder2025retraceinteractivevisualizationsreasoning}; InconLens focuses on repeated run diagnosis through semantically meaningful milestones~\cite{yan2026inconlensinteractivevisualdiagnosis}; and SeaView and related visual analytics work compare software engineering agent runs across models, tasks, and workflow settings~\cite{bula2025seaviewsoftwareengineeringagent, wang2025illuminatingllmcodingagents}.

Together, these systems show the growing need for tools that structure, summarize, and support interactive analysis of long agent traces. 
%These systems show that raw logs are a poor interface for complex agent behavior. 
However, no existing tool supports the fine-grained analysis of APR trajectories, where users label action categories and semantic relations among TAR elements, then inspect how those labels reveal alignment, repetition, divergence, ignored feedback, and other process-level patterns within a repair attempt.

We present \textit{TraceView}, an interactive tool for labeling and visualizing software engineering agent trajectories. TraceView represents each run as a directed graph over \textit{Thought}, \textit{Action}, and \textit{Result} nodes. Users can upload raw agent trajectories, assign action categories, label semantic relations, export the labels into a reusable viewer format, and inspect the resulting graph. They can also analyze pre-labeled trajectories from the bundled dataset. TraceView provides two coordinated graph modes. The iteration view summarizes the run at the step level, while the detailed view expands each step into separate TAR nodes. Furthermore, TraceView provides relation filters, legends, relationship metrics, patch outcome summaries, and an inspector panel that helps users isolate productive flow, repeated behavior, negative relations, or no influence edges for each node.

We evaluated TraceView through a survey-based user study with five researchers. After using the tool, participants answered Likert-scale and open-ended questions about its perceived usefulness, clarity, evidence availability, and whether moving from trajectory overviews to detailed inspection helps them better understand APR agent behavior.

This paper makes the following contributions:

\begin{itemize}[leftmargin=*, nosep]
    \item We present \textit{TraceView}, an interactive tool for labeling and visualizing APR agent trajectories as structured \textit{Thought}, \textit{Action}, and \textit{Result} graphs.
    \item We provide a relation labeling and analysis workflow with semantic edges, relation filters, patch results, relationship metrics, and node-level evidence inspection.
    \item We report a preliminary user study showing that users found TraceView useful for scanning trajectories and understanding repair workflows, while identifying concrete improvements for graph readability and navigation.
\end{itemize}

\section{TraceView}

\begin{figure*}[t]
    \centering
    \begin{minipage}[t]{0.49\textwidth}
        \centering
        \includegraphics[width=\linewidth]{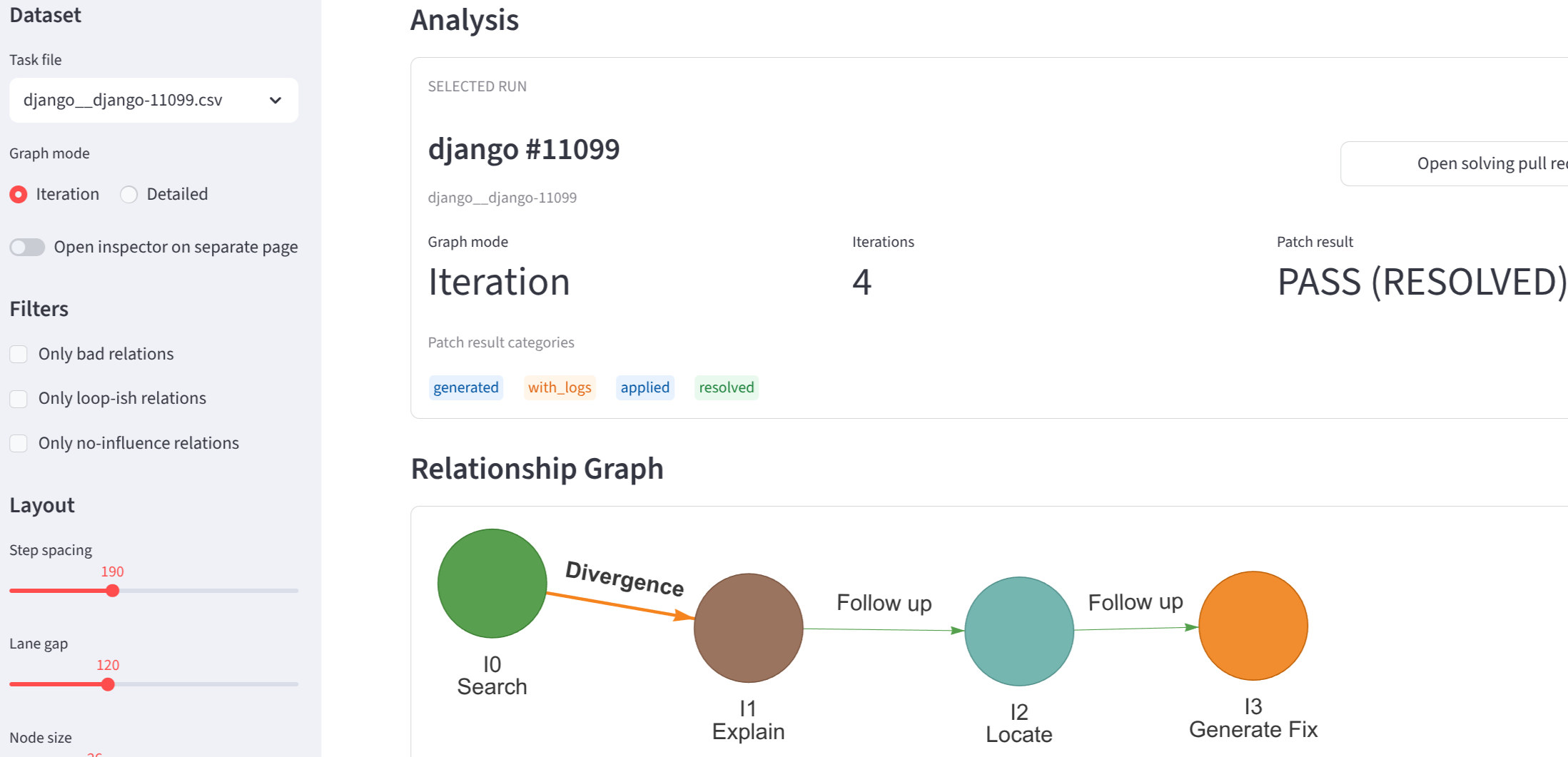}\\[-0.5ex]
        {\footnotesize (a) Iteration view.}
    \end{minipage}
    \hfill
    \begin{minipage}[t]{0.49\textwidth}
        \centering
        \includegraphics[width=\linewidth]{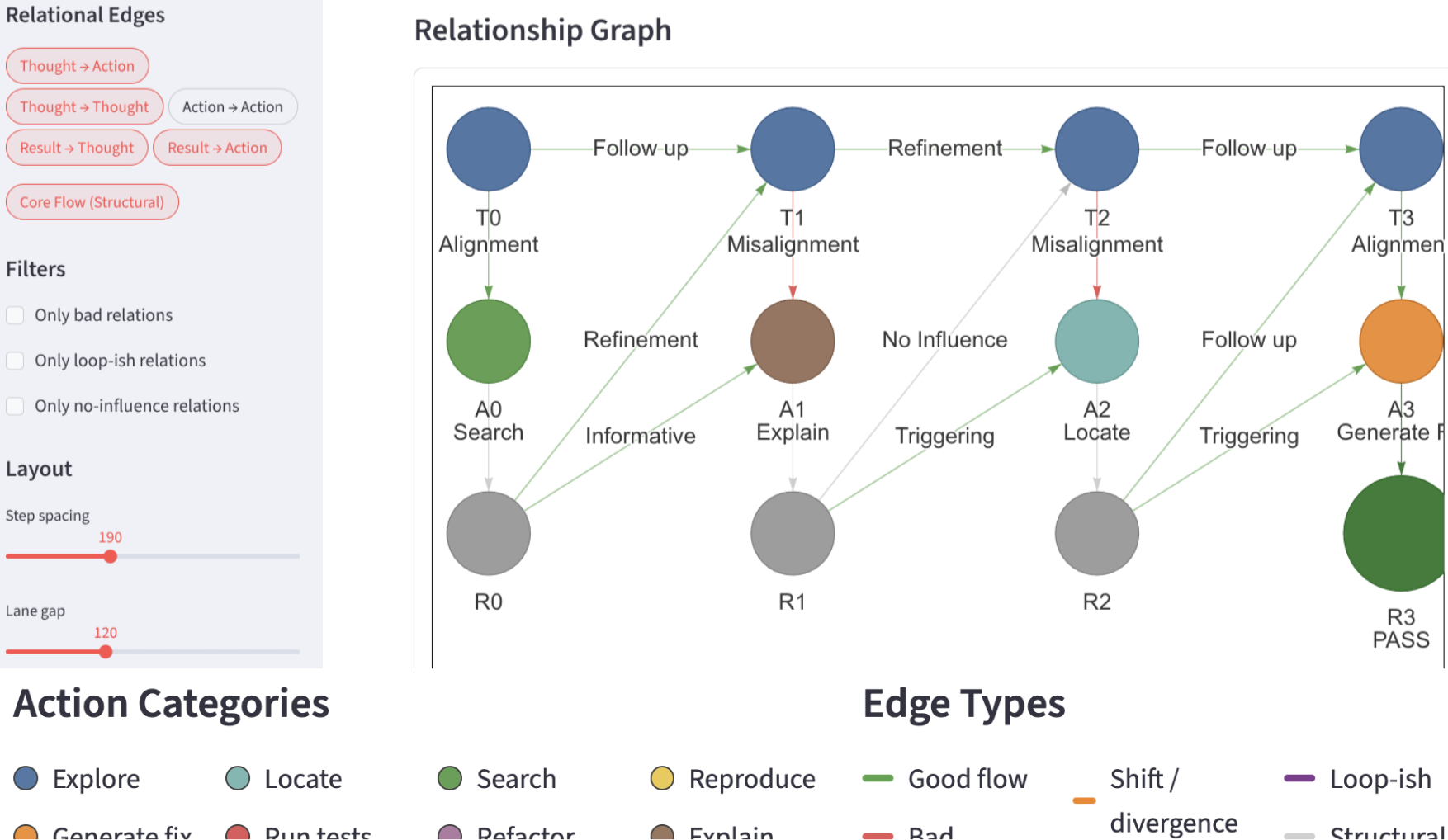}\\[-0.5ex]
        {\footnotesize (b) Detailed TAR view.}
    \end{minipage}
    \caption{TraceView provides two graph views for APR trajectory analysis. The iteration view summarizes each step by action category, while the detailed view expands each step into \textit{Thought}, \textit{Action}, and \textit{Result} nodes with semantic relation edges.}
    \label{fig:traceview-analysis-modes}
    \vspace{-1.5em}
\end{figure*}

TraceView is a tool designed for researchers and developers who need to inspect and annotate APR agent behavior beyond final patch success or failure. TraceView makes the TAR structure of raw logs visible, highlights semantic relationships between trajectory elements, and lets users move between overview, filtering, and details on demand during trajectory analysis~\cite{545307}.

\subsection{Workflow and Architecture}

TraceView is implemented as a Streamlit application with three user-facing sections. The \textit{Labeling} section ingests uploaded trajectories, presents action labeling and relation labeling tasks, and exports labels into the same CSV-based structure used by the viewer. The \textit{Overview} section lists available bundled and user-exported runs. The \textit{Analysis} section renders the selected run as an interactive graph and provides legends, relation metrics, patch result summaries, graph controls, and an inspector for the selected node.

Internally, TraceView keeps raw trajectory content, user labels, and graph state separate. Action labels are stored at the step level, while semantic relation labels are stored at the edge level, allowing labeled runs to be reopened, analyzed, and reused without reprocessing the original trace.

\subsection{Input Artifacts and Label Vocabulary}

TraceView accepts both raw and pre-labeled inputs. Raw inputs include SWE agent style JSON trajectories~\cite{yang2024sweagent}, JSONL command traces, plain text or log files with \textit{Thought}, \textit{Action}, and \textit{Result} sections, and zip archives containing multiple supported trajectory files. For bundled data, TraceView loads AutoCodeRover style APR trajectories~\cite{zhang2024autocoderover}, reconstructed text logs, action category CSV files, relation label CSV files, and patch outcome metadata.

Supported inputs are normalized into ordered \textit{Thought}, \textit{Action}, and \textit{Result} steps before labeling or analysis. These normalized steps give both workflows the same underlying representation, whether the run comes from an uploaded raw trace or a bundled pre-labeled trajectory.

Following prior work on TAR trajectory analysis of software engineering agents~\cite{bouzenia2025understandingsoftwareengineeringagents}, TraceView supports the action categories and semantic relation labels shown in Table~\ref{tab:traceview-labels}. Action categories describe what the agent does at each step, while relation labels describe how one TAR component relates to another. Relation labels define which types of TAR components are being connected, such as a thought to an action within the same step or a result to a consecutive thought. %These labels help users describe whether a run follows evidence, repeats prior behavior, changes direction, or misunderstands feedback.
By attaching a semantic label to each source--target TAR relation, TraceView lets users characterize how the repair process evolves. For example, if a test result reports a failing assertion and the next thought uses that failure to plan a fix, the Result$\rightarrow$Thought edge can be labeled \emph{Informative}; if the next thought instead draws an incorrect conclusion from the same feedback, that edge can be labeled \emph{Misinterpretation}.

\begin{table}[t]
\centering
\caption{Action categories and relation labels supported by TraceView.}
\label{tab:traceview-labels}
\small
\begin{tabular}{p{0.27\linewidth}p{0.63\linewidth}}
\hline
\textbf{Label type} & \textbf{Labels} \\
\hline
Action categories &
\textit{Explore}, \textit{Locate}, \textit{Search}, \textit{Reproduce}, \textit{Generate fix}, \textit{Run tests}, \textit{Refactor}, \textit{Explain} \\
\hline
Relation labels &
\textit{Alignment}, \textit{Follow-up}, \textit{Refinement}, \textit{Redundancy}, \textit{Repetition}, \textit{Divergence}, \textit{Contradiction}, \textit{Informative}, \textit{Triggering}, \textit{No influence}, \textit{Misalignment}, \textit{Misinterpretation} \\
\hline
\end{tabular}
\vspace{-1.5em}
\end{table}

\subsection{Labeling Workflow}

\begin{figure*}[t]
    \centering
    \includegraphics[scale=0.32]{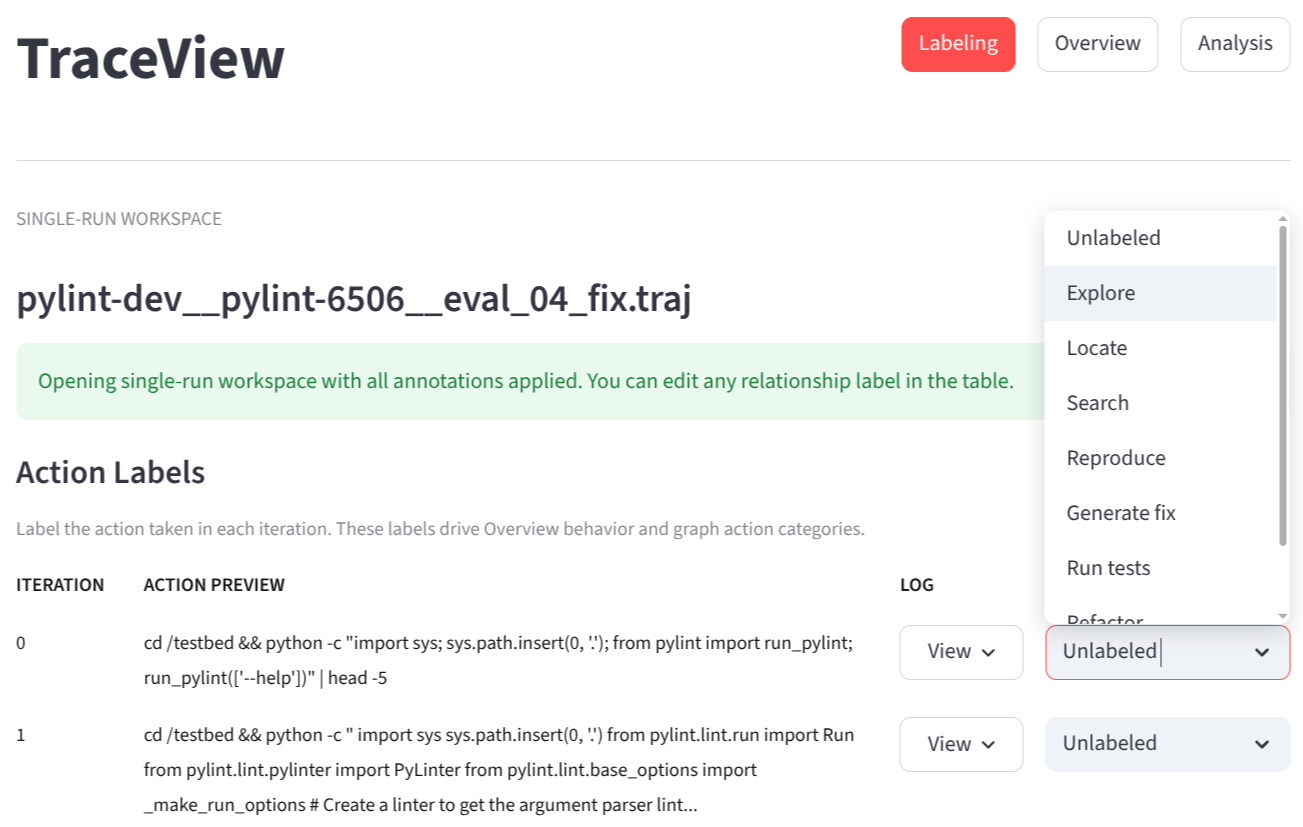}
    \caption{TraceView's labeling workspace lets users assign action categories and inspect source evidence before labeling semantic relations.}
    \label{fig:traceviewlabeling}
    \vspace{-.5cm}
\end{figure*}

Figure~\ref{fig:traceviewlabeling} shows the labeling workspace. The labeling workflow follows the dependency structure of the data. Users label actions first because action categories are needed to gather iteration-level context. They then label semantic relations between source and target components. Before relation labeling, TraceView creates relation candidates from fixed source--target templates. A candidate at step $i$ may connect nodes within the same step or across adjacent steps, depending on the relation template.

For each relation candidate, the interface shows the relation to be labeled, the relevant steps, the evidence text from both TAR components, and the available labels. For example, when labeling whether a test result influences the agent's next thought, the user can inspect the result text and the following thought before selecting a label such as \textit{Informative}, \textit{No influence}, or \textit{Misinterpretation}. Longer evidence can be opened in a popover, reducing table clutter while preserving access to the original text.

Completed labels can be exported as JSON or viewer-compatible CSV files for graph analysis. When a user sends a labeled trajectory to the viewer, TraceView writes the action categories, relation files, reconstructed TAR log, and metadata for the Overview and Analysis sections.

\subsection{Graph Views}

Figure~\ref{fig:traceview-analysis-modes} shows the two graph views supported by TraceView. The iteration view provides a compact entry point for scanning the run at the step level, while the detailed view expands each step into separate \textit{Thought}, \textit{Action}, and \textit{Result} nodes.

TraceView transforms raw agent logs into a graph-based intermediate representation before rendering~\cite{faust2024anteaterinteractivevisualizationprogram, lu2024agentlensvisualanalysisagent, 10.1609/aaai.v39i28.35350}. In detailed mode, each source step $i$ is expanded into three nodes, $T_i$ for the thought, $A_i$ for the action, and $R_i$ for the result. Thought and result nodes use fixed role colors, while action nodes are colored by their action category. For bundled trajectories, patch outcome metadata is reduced to a primary status such as \textit{RESOLVED}, \textit{APPLIED}, \textit{GENERATED}, or \textit{UNKNOWN}. This status determines the terminal result node annotation as \textit{PASS}, \textit{FAIL}, or \textit{UNSCORED}.

The semantic edges are constructed from five relation families. These families connect \textit{Thought$\rightarrow$Action} nodes within the same step, \textit{Thought$\rightarrow$Thought} nodes across adjacent steps, \textit{Action$\rightarrow$Action} nodes across adjacent steps, \textit{Result$\rightarrow$Thought} nodes across adjacent steps, and \textit{Result$\rightarrow$Action} nodes across adjacent steps. TraceView can also render structural flow edges to preserve the local TAR sequence and chronological progression when semantic filters are disabled.

In iteration mode, TraceView renders one node per source iteration, labeled by iteration number and action category, with optional context derived from the reconstructed log. Action-to-action relations are displayed between iteration nodes, and structural chronological edges are added when no relation filter is active. Users can switch to detailed mode when they need the full TAR decomposition.

\subsection{Filtering, Inspection, and Metrics}

TraceView maps relation labels into visual families. Productive continuation labels such as \textit{Alignment}, \textit{Follow-up}, \textit{Refinement}, \textit{Informative}, and \textit{Triggering} are rendered as flow edges. \textit{Divergence} is shown as a direction shift, \textit{Redundancy} and \textit{Repetition} as iterative loops, \textit{Misalignment}, \textit{Misinterpretation}, and \textit{Contradiction} as negative relations, and \textit{No influence} is treated as a neutral relation. This grouping allows users to filter the graph by relation types, making it easier to isolate productive or unproductive flows, consistent with interactive behavioral analysis tools that help users isolate meaningful subsets rather than inspect a single undifferentiated view~\cite{Cabrera2023Zeno}.

Selecting a node opens an inspector with the underlying thought, action, or result text and the labeled relations connected to that node. This interaction follows details on demand principles and aligns with explanatory and interactive debugging work that emphasizes user guided diagnosis over passive display~\cite{545307, 10.1145/2678025.2701399, Epperson_2025, hutter2026agentstepperinteractivedebuggingsoftware}. TraceView also reports aggregate relation counts by label and edge family, giving users a compact summary of the dominant patterns in the selected trajectory.

\section{Preliminary Evaluation}

We conducted a preliminary user study to assess whether TraceView is understandable and useful for researchers or developers inspecting APR agent trajectories. Because the goal of the study was design feedback rather than statistical validation, we focused on perceived clarity, evidence availability, workflow usefulness, and open-ended comments about confusing or missing interface behavior.

\subsection{Participants and Procedure}

We recruited five computer science researchers to participate in the user study. Participants first used TraceView with an APR trajectory sample and followed the workflow from labeling to graph analysis. The task asked them to inspect the repair process, use the available views, and identify confusing or problematic elements of the UI when possible. The participants completed a survey that included both five-point Likert items and open-ended questions.

The survey covered three parts of the interface: 1) the \textit{labeling workflow}, including whether the main sections and accepted input formats were clear, whether action previews were long enough for labeling decisions, whether longer-log popovers were useful, and whether the labeling table was dense without being overwhelming; 2) the \textit{relationship evidence and graph interpretation}, including whether row-level popovers provided enough source and target evidence and whether the graph could be interpreted without guessing; and 3) the \textit{analysis workflow}, asking whether TraceView was more efficient than raw logs, whether the graph made trajectories easier to scan, and whether the overview/detail hierarchy supported analysis and orientation, with each part also including open-ended questions about confusing renderings, missing information, or interface suggestions. The survey form and study responses are included in our replication package. 

\subsection{Survey Results}

Table~\ref{tab:survey-results} summarizes the survey responses across the three parts of the study. Overall, participants rated TraceView positively across the labeling, relationship-evidence, and analysis workflows. For the labeling workflow, participants rated the clarity of the main sections (median = 4) and rated the accepted input formats and longer log popovers highly (median = 5). For relationship evidence and graph interpretation, row-level evidence popovers received a median of 5, while graph interpretability received a median of 4. For the analysis workflow, participants rated TraceView as more efficient than raw logs, the graph easier to scan, and the overview/detail view useful for analysis, all with medians of 5.

The lower minimum scores for action-preview length and graph interpretability show that some participants needed more concise and informative cues. Open-ended comments suggested that short previews sometimes lacked enough context, while longer previews could be difficult to scan. 

\begin{table}[t]
      \centering
      \caption{Selected Likert results from the formative evaluation ($n=5$).}
      \vspace{-0.2cm}
      \label{tab:survey-results}
      \begin{tabular}{p{0.62\columnwidth}ccc}
        \hline
            Survey item & Min & Median & Max \\
        \hline
            \multicolumn{4}{l}{\textit{\textbf{Labeling workflow}}} \\
            Main sections are easy to understand & 3 & 4 & 5 \\
            Accepted input formats are clear & 4 & 5 & 5 \\
            Action previews are long enough for labeling & 1 & 4 & 5 \\
            View popover is useful for longer logs & 4 & 5 & 5 \\
        \hline
            \multicolumn{4}{l}{\textit{\textbf{Relationship evidence and graph interpretation}}} \\
            Row-level popovers provide enough evidence & 3 & 5 & 5 \\
            Graph is interpretable without guessing & 1 & 4 & 5 \\
        \hline
            \multicolumn{4}{l}{\textit{\textbf{Analysis workflow}}} \\
            TraceView is more efficient than raw logs & 4 & 5 & 5 \\
            Graph makes trajectory easier to scan & 4 & 5 & 5 \\
            Overview/detail view is useful for analysis & 3 & 5 & 5 \\
        \hline
      \end{tabular}
  \end{table}

\subsection{Feedback and Suggestions for Future Improvement}

Open-ended feedback identified several concrete strengths and limitations. Participants found the graph useful for understanding the repair process, but wanted node labels to carry short summaries rather than only identifiers e.g., $T0$ or $A0$. Some comments requested hover previews or concise summaries for TAR nodes. Participants also noted cases where diagonal edges or source--target relations were difficult to interpret without additional explanation, especially when source previews were empty or when relation labels such as \textit{Misalignment} were not immediately obvious from the graph.

These findings are useful as they point to changes that can be made without altering the core data model. Addition of short node summaries and hover previews would make the current graph easier to interpret while preserving the TAR representation and semantic relation workflow.

\section{Conclusions and Future Work}

TraceView can help researchers and developers inspect APR agent behavior as a process rather than as only a final patch outcome. It parses trajectories into TAR components, supports action and relation labeling, and renders the resulting trace as an interactive graph with filters, metrics, patch-status summaries, and node-level evidence. Together, these features aim to make it easier to study where an agent follows useful feedback, repeats the same actions, diverges from the task, or misinterprets the repair context.

Our preliminary user study suggests that the participants found TraceView to improve trajectory scanning compared with raw logs and that the overview-to-detail workflow is useful for understanding repair runs. The study also identified clear next steps, including the addition of shorter node summaries and hover previews. In our future work, we plan to incorporate these improvements and conduct a larger user study with longer and more diverse tool trajectories to measure the tool's usefulness for identifying common success and failure patterns in agentic APR workflows.

\bibliographystyle{IEEEtran}
\bibliography{ref}

\end{document}